\documentclass[12pt,a4paper]{article}
\pdfoutput=1

 \usepackage{jheppub}
 \usepackage{amsmath, amssymb, latexsym, amsthm, amsfonts, mathrsfs, amscd, mathtools,longtable}
 \usepackage{caption}
 \usepackage{subcaption}
 \usepackage{slashed}  
\usepackage{float}
\usepackage{graphicx}
\graphicspath{ {project2 tex/} }

\def\m{\mu}

\def\k{\kappa}

\newcommand{\be} {\begin{equation}}
\newcommand{\ee} {\end{equation}}
\newcommand{\bea} {\begin{eqnarray}}
\newcommand{\eea} {\end{eqnarray}}
\newcommand{\ba} {\begin{array}}
\newcommand{\ea} {\end{array}}
\newcommand{\nn} {\nonumber}

\theoremstyle{plain}

\theoremstyle{definition}

 \title{Numerical experiments on coefficients of instanton partition functions}
 \author{Aradhita Chattopadhyaya$^{1}$, Jan Manschot$^{2,3,4}$}
 \affiliation{$^1$School of Theoretical Physics, Dublin Institute of Advanced Studies, Dublin 4, Ireland
   $^2$School of Mathematics, Trinity College, Dublin 2,
   Ireland\\
   $^3$Hamilton Mathematical Institute, Trinity College, Dublin 2,
   Ireland\\
   $^4$School of Natural Sciences, Institute for Advanced Study, 1
  Einstein Drive, Princeton, NJ 08540 USA}
  
\emailAdd{aradhita@stp.dias.ie, manschot@maths.tcd.ie}
 
\abstract{We analyze the coefficients of partition functions of
  Vafa-Witten theory for the complex projective plane
  $\mathbb{CP}^2$. We experimentally study the growth of the coefficients for
  gauge group $SU(2)$ and $SU(3)$, which are examples of mock modular
  forms of depth $1$ and 2 respectively. We also introduce the notion of ``mock cusp
  form'', and study an example of weight 3 related to the $SU(3)$ partition function.
  Numerical experiments on the first 200 coefficients suggest that the
  coefficients of a mock modular form of weight $k$ grow as the
  coefficients of a modular form of weight $k$, that is to say as
  $n^{k-1}$. On the other hand the coefficients of the mock cusp form
  appear to grow as $n^{3/2}$, which exceeds the growth of classical
  cusp forms of weight 3. We
  provide bounds using saddle point analysis, which however largely exceed the
  experimental observation. 
  \\
\today
}

\begin{document}
\maketitle

\section{Introduction}
 
Instantons are (anti-)self-dual solutions of the Yang-Mills equations of
motion. These non-perturbative solutions are central in theoretical
physics \cite{tHooft:1974kcl, Belavin1975, Losev:1997tp, Dorey:2002ik,
  Nekrasov:2002qd, Vandoren:2008xg}, and
connect to many subjects including geometry \cite{Uhlenbeck1986, DONALDSON1990, Donaldson90,
  Klyachko:91, Yoshioka94, Kool_2014} and analytic number
theory \cite{Gottsche1990, Vafa:1994tf,
  Gottsche:1996aoa, Bringmann:2010sd, Coskun2022}. Of particular interest are moduli spaces of instanton solutions, and their
topological invariants. This article considers generating functions of
Euler numbers of instanton moduli spaces for the four-manifold
$\mathbb{P}^2$. Rather strikingly, these partition functions are
examples of (mock) modular forms as a consequence of electric-magnetic
duality \cite{MONTONEN1977117, Vafa:1994tf, Verlinde:1995mz}, and even give  
rise to new functions of this type \cite{Manschot:2010nc,
  Manschot:2014cca, Alexandrov:2016enp, Manschot:2017xcr}. 
     
Physically, these generating functions are the partition functions of
a specific topological twist of $\mathcal{N}=4$ Yang-Mills theory
\cite{Witten:1988ze,Vafa:1994tf} on a four-manifold $X$. The supersymmetry makes an explicit analysis feasible
for many physical quantities of interest. If $X$ is Ricci flat, for
example $X=K3$, the partition functions of this theory are
known to give rise to classical modular forms. On other four-manifolds, such as the rational
surfaces, the partition functions are instances of mock modular forms
\cite{Vafa:1994tf, Manschot:2011dj, Dabholkar:2020fde}, and even mock modular forms of higher depth
\cite{Alexandrov:2016enp, Manschot:2017xcr}. More precisely, the 
functions are examples of mixed mock modular forms, and their
coefficients grow exponentially. An exact formula of Rademacher type
was derived for these coefficients for gauge group $SU(2)$ in \cite{Bringmann:2010sd},
and gauge group $SU(3)$ in \cite{Bringmann:2018cov}.  
  
This article concerns the factor of the partition function which
are pure mock modular forms. These are attributed to the
smooth instanton solutions, or the locally free sheaves. For gauge group
$SU(2)$ these coefficients are famously Hurwitz class numbers
\cite{Klyachko:91, zagier:75}. 
This article explores the coefficients for gauge group $SU(3)$. We
find intriguing patterns for the coefficients of these partition
functions, while we furthermore determine an upperbound on the asymptotic
growth of these coefficients. In our analysis, we introduce the notion
of a ``mock cusp form''. For our specific example of a mock cusp form
of weight $k=3$, we find experimentally that the coefficients of the
function accidentally grow as $\sim n^{3/2}$. We are unable to prove
this growth. Using the saddle point method, we are able to bind the
growth of the coefficients by $n^{5/2}$, and using
a more heuristic argument based on lattice sums by $n^2$.
To put these bounds into context, we note that the saddle point method
applied to classical cusp forms of weight $k$ (or Hecke bound) gives
$n^{k/2}$, which is notably larger than the sharper
Deligne-Petersson-Ramanujan bound $n^{(k-1)/2}$. It would be
interesting to explore the growth of similar functions, such as the coefficients of
VW partition functions for $N>3$, and those of generating functions of
bound states of black holes \cite{Chattopadhyaya:2021rdi}.  

The outline of this article is as follows. We briefly review VW theory
in section \ref{sec21}, followed
by a discussion of the partition functions for gauge groups $SU(2)$ and $SU(3)$ in section \ref{sec22}. In section \ref{sec31}, we
define modular forms and their transformation properties. In section
\ref{sec32}, we define the mock modular forms, state their
transformation properties and extend the definition to mock cusp
forms. Section \ref{sec5} provides detailed numerical results and
plots obtained for the partition functions of $SU(3)$  VW theory and
how these functions grow for $p$-th coefficients for prime $p$.
Finally in section \ref{sec4}, we analyse the asymptotics of the
Fourier coefficients of the partition functions of VW theory for gauge
group $SU(3)$. In section  \ref{sec41}, we review the rough bound on
the growth of coefficients of modular and cusp forms. We extend this
in section \ref{sec42} to the growth of coefficients of mock cusp
forms associated with the partition functions of VW theory. Section
\ref{sec53} provides an heuristic argument for the growth of the
coefficients of theta series.

\section{Yang-Mills theory and mock modularity}\label{sec2}

This section briefly reviews ${\cal N}=4$ super Yang-Mills theory and
its topologically twisted version, Vafa-Witten theory
\cite{Vafa:1994tf}. For more detailed expositions in mathematics and
physics we suggest
\cite{Klyachko:91,Nakajima:1993jg,Nakajima:1994nid,Yoshioka94,Yoshioka:k95},
and \cite{Dorey:1996hu,Dorey:1996bf,Dorey:1999pd, Dorey:2002ik}.

\subsection{Instanton solutions}\label{sec21}
For a simply connected gauge group $G$, we let $F_{ij}$ be the field strength for the gauge potential $A_i$ with $i,j\in\{1,\dots,4\}$,
\begin{eqnarray} 
F_{ij} &=& \partial_i A_j-\partial_j A_i+[A_i,A_j].
\end{eqnarray}
The Yang-Mills action reads
\be
\label{action}
S = \frac{1}{4g^2}\int_X  d^4x\, {\rm Tr}\;F_{ij}F^{ij},
\ee
where the trace is over the indices of the representation of the Lie
algebra $G$. The action $S$ is left invariant by the gauge symmetry, whic acts on the covariant derivative $D_i=\partial_i
+A_i$ as,
\begin{eqnarray}
D_i\rightarrow h^{-1}D_i h,\; {\rm for}\;
  h(\vec x)\in  G.  
\end{eqnarray}
The action is bounded below by the instanton number
 \be
k=-\frac{1}{8\pi^2}\int_X {\rm Tr}\;F\wedge F,
\ee
which is a topological invariant of a solution to the Yang-Mills equations of motion.
The instanton number gives a lower bound on the action,
\begin{eqnarray}
  \label{Sbound}
S &=& \frac{1}{4g^2}\int_X  d^4x\, {\rm Tr}\;F_{ij}F^{ij}\\ \nn
&=& \frac{1}{8g^2}\int_X  d^4x\, {\rm Tr}\;\left((F^{\pm})^2\mp  2F\wedge F\right)\\ \nn
&\ge & \pm \frac{2\pi^2}{g^2}\,k,
\end{eqnarray}
where $F^\pm$ are the self dual and anti-self dual decomposition of $F_{ij}$:
\begin{eqnarray}
2F_{ij}^+ &=& F_{ij}+\frac{1}{2}\epsilon_{ijkl}F^{kl},\quad 2F_{ij}^- = F_{ij}-\frac{1}{2}\epsilon_{ijkl}F^{kl}.
\end{eqnarray}
Solutions which saturate the bound (\ref{Sbound}), those with with $F^+=0$ or $F^-=0$, are called ``instantons'' or ``anti-instantons''.

Instanton solutions were studied by several physicists as well as
mathematicians since 1970s. On the physics side, instantons are related to
magnetic monopole solutions by dimensional reduction
\cite{tHooft:1974kcl,tHooft:1976snw,tHooft:1976rip,tHooft:1986ooh}. A complete construction using linear algebra alone
for the Yang-Mills self dual instantons in Euclidean $S^4$ was first
given in \cite{Atiyah:1978ri,Drinfeld:1978xr}.
 
Modulo gauge transformations, the moduli space $\mathcal{M}_k$ of instanton solutions
with instanton number $k$ is finite dimensional. The dimension
corresponds physically to the number of fermionic zero modes (ground
states), or more precisely the index of the Dirac
operator. The sequence of topological invariants of instanton moduli
spaces as function of the Euler number is naturally combined to a generating function. 
For example, the generating function of Euler numbers of moduli spaces
of instantons on $\mathbb{R}^4$
with boundary conditions for the gauge potential, $A|_{r\rightarrow \infty}\sim \frac{1}{r^2}$ and
$F|_{r\rightarrow \infty}\sim \frac{1}{r^3}$ for $r\to \infty$ were
computed in \cite{Nakajima:1993jg,Nakajima:1994nid}. Such generating
functions are often realized physically as a statistical partition
function of a topologically twisted supersymmetric Yang-Mills theory. Depending on the field
content, different topological observables can be realized \cite{Losev:1997tp}. The
subject of this paper is the Vafa-Witten twist of $\mathcal{N}=4$
Yang-Mills theory, whose partition function is a generating function
of Euler numbers of instanton moduli spaces.

We recall a few aspects from \cite{Vafa:1994tf} in what follows.
The action of the bosonic fields of ${\cal N}=4$ super Yang-Mills
theory reads 
\be
\begin{split}
  &S_b(A_i,v_a) =\\
  &\quad \frac{1}{2g^2}\int_X  d^4x\, {\rm
  Tr}\;\left(\frac{1}{2}F_{ij}F^{ij}+\sum_{a,b=1}^6[v_a,v_b]+\sum_{a=1}^6
  (D_iv_a)^2 +\frac{ig^2\theta}{4\pi^2} F\wedge F\right),
\end{split}
\ee 
where $v_a$, $a=1,\dots, 6$, are scalar fields, and $g$ is the gauge
coupling, $X$ is a 4-manifold. The theory has a global $R$-symmetry
group $SU(4)$. The four scalars $v_a$ transform under the
6-dimensional representation of $SU(4)$, whereas the gauge field
$A_i$ is a singlet. The four supercharges transform under the
four-dimensional representation ${\bf
  4}$ of $SU(4)$.

A topological twist of this $\mathcal{N}=4$ theory on a compact four-manifold $X$ identifies a
principal $SU(4)$ R-symmetry bundle. The identification follows by the
action on the associated bundle with fiber the four-dimensional
representation ${\bf 4}$ of $SU(4)$. The Vafa-Witten twist is the
twist for which the ${\bf 4}$ is identified with the
representation $({\bf 1}, {\bf 2})\oplus ({\bf 1}, {\bf 2})$ of the
local frame group $Spin(4) \simeq SU(2)_+\times SU(2)_-$. The
identification of the principal $SU(4)$ R-symmetry bundle thus
follows from the chiral spin bundle $W^-$ on $X$, i.e. the bundle
associated to the $({\bf 1}, {\bf 2})$ representation of $Spin(4)$. Crucially, the
supercharges of the topologically twisted theory includes two scalar
supercharges. The fermionic fields of the theory consists of two self-dual
two-forms, two vectors and two scalars, whereas the bosonic field content of the twisted theory is:
\begin{enumerate}
\item Gauge field $A_i$,
\item Self-dual two-form $B_{ij}$. For $i=0$, we abbreviate $B_{0j}=B_j$,
\item Real scalar $C$, and complex scalar $\phi$.
\end{enumerate}
The bosonic part of the action with $\phi,\bar{\phi}$ set to zero is given by,
\be
\begin{split}
   S(A_i, B_i, C) & =\frac{1}{2g^2}\int_X\,  d^4x\, {\rm
  Tr}\left(F_{ij}^2+ (D_iB_j)^2+ (D_i
  C)^2 \right. \\
&\quad \left.   +\sum_{i<j}[B_i,B_j]^2+\sum_i [C,B_i]^2
  +\frac{ig^2\theta}{4\pi^2} F\wedge F\right).
\end{split}
\ee
The instanton solutions are given by
\begin{eqnarray}
B_{ij}=C=F_{ij}^+=0 \quad {\forall}\; i,j.
\end{eqnarray}
In addition, there is a monopole or Abelian branch \cite{Vafa:1994tf}. This latter branch
is absent on manifolds with positive scalar curvature such as
$\mathbb{CP}^2$.

\subsection{Partition functions of Vafa-Witten theory}\label{sec22}
We review the partition functions of the Vafa-Witten theory on gauge groups $SU(N)$ for the four manifolds $K3$ and $\mathbb{CP}^2$.
The partition function is given schematically by the path integral
formalism as
\begin{eqnarray} 
Z(\tau,\bar \tau) &=& \int {\cal D}\vec\Phi\, e^{-S(\vec\Phi)},
\end{eqnarray} 
where $\vec\Phi$ represents the field content of the theory, which includes the fermionic super-partners, ghost anti-ghost pairs and auxiliary fields and
\begin{eqnarray}
\tau=\frac{\theta}{2\pi}+\frac{4\pi i}{g^2},
\end{eqnarray}
is the complexified coupling constant. 

The partition functions of Vafa-Witten theory on $K3$ and gauge group  $U(1)$ were given in \cite{Vafa:1994tf}
\be
  \label{ZK3}
Z^{K3}_1(\tau) = \frac{1}{\eta^{24}(\tau)}=q^{-1}\sum_{k\geq 0}p_{24}(n)q^k,
\ee
where $q=e^{2\pi i\tau}$, $p_{24}(k)$ is the partition of a  positive integer $k$ in 24
colors, and $\eta(\tau)$ is the Dedekind eta function given by,
\be
\label{Deta}
\eta(\tau)=q^{1/24}\prod_{n=1}^{\infty}(1-q^n). 
\ee 
The function in (\ref{ZK3}) is also the partition function of 24 free
scalar fields in two dimensions.
In the twisted $SU(2)$ theory the partition function is given by,
\begin{eqnarray} 
Z^{K3}_{2}(\tau) &=& \frac{1}{8}Z_{K3}(2\tau)+\frac{1}{4}Z_{K3}(\frac{\tau}{2})+\frac{1}{4}Z_{K3}(\frac{\tau+1}{2})
\end{eqnarray} 
More generally, for gauge group $SU(N)$ with prime $N$ the result was
proposed in \cite{Vafa:1994tf}, 
\begin{eqnarray}
Z^{K3}_{N} &=& \frac{1}{2N^2}Z_{K3}(N\tau)+\frac{1}{2N}\sum_{b=0}^{N-1}Z_{K3}(\frac{\tau+b}{N}) .
\end{eqnarray}
The partition function for other 't Hooft fluxes can be expressed in a
similar form \cite{Vafa:1994tf}. Explicit partition functions for other
algebraic surfaces with $b_2^+>1$ are expressed in terms of
Seiberg-Witten invariants \cite{Dijkgraaf:1997ce, Tanaka:2017jom, Gottsche:2017vxs}.

We continue with the four-manifold $\mathbb{CP}^2$. While the
partition functions for K3 are given in terms of modular forms for any
$N$, we will see that the partition functions for $\mathbb{CP}^2$ give
rise to new functions. The result for gauge group $U(1)$ is still
given in terms of the Dedekind eta function, 
\begin{eqnarray}
Z^{\mathbb{CP}^2}_{U(1)} &=& \frac{1}{\eta^3(\tau)}.
\end{eqnarray}
 
However for $SU(2)$ with 't Hooft flux $\mu$, the partition functions read \cite{Vafa:1994tf}:
\be\label{pfsu2}
Z^{\mathbb{CP}^2}_{2,\mu}(\tau,\bar \tau) = \frac{\widehat
                                              f_{2,\mu}(\tau,\bar
                                              \tau)}{\eta^6(\tau)}, \qquad \mu=0,1,
\ee
with the $\widehat f_{2,\mu}$, explicitly given as
\begin{eqnarray}
  \label{hatf2mu}
\widehat f_{2,\mu}(\tau,\bar{\tau}) &=& f_{2,\mu}(\tau)-\frac{3i}{4\sqrt{2}\pi}\int_{-\bar{\tau}}^{i\infty} \frac{\Theta_{\mu/2}(w)}{(-i(w+\tau))^{3/2}}dw.
\end{eqnarray}
The holomorphic parts $f_{2,\mu}$ are a multiple of the generating
functions $G_\mu$ of Hurwitz class
numbers $H(n)$ \cite{Yoshioka94,Yoshioka:k95,Klyachko:91,zagier2, zagier:75},
\be
  \label{f2mu}
f_{2,\mu}(\tau) = 3G_\mu(\tau),\qquad
G_\mu(\tau)=\sum_{n=0}^{\infty}H(4n-\mu)\,q^{n-\mu/4}.
\ee
Moreover, $\Theta_{\mu/2}$ in (\ref{hatf2mu}) is the theta series defined by
\be
\label{Thetamu}
\Theta_{\alpha}(\tau)= \sum_{n\in\mathbb{Z}+\alpha}q^{n^2/2}.
\ee
See equation (\ref{f2021series}) for explicit generating series for the
$f_{2,\mu}(\tau)$. Table \ref{coefclass} in Section \ref{sec5} lists
the first few Hurwitz class numbers.

The holomorphic part $f_{2,\mu}$ transforms as
\be
\begin{split}
f_{2,\mu}\!\left(-\frac{1}{\tau}\right) &=
                               -\frac{\tau^{3/2}}{\sqrt{2}}\sum_{\nu=0}^1 (-1)^{\mu\nu} \left(f_{2,\nu}(\tau) -\frac{3i}{4\sqrt{2}\pi} \int_{0}^{i\infty}\frac{\Theta_{\nu/2}(w)}{(-i(w+\tau))^{3/2}}\;dw\right),\\
f_{2,\mu}(\tau+1)&=  e^{-\pi i \mu^2/2} f_{2,\mu}(\tau).
\end{split}
\ee
For the non-holomorphic function $\widehat f_{2,\mu}$, the shift by 
the period integral is absorbed by the non-holomorphic
integral such that $\widehat f_{2,\mu}$ transforms as a modular form
of weight 3/2. The full partition function is thus a non-holomorphic modular form of
weight $-3/2$. Physically one can derive the holomorphic anomaly using
localization techniques on the Coulomb branch of the effective field
theory \cite{Dabholkar:2020fde, Bonelli:2020xps, Manschot:2021qqe}.

For $X=\mathbb{CP}^2$, the partition functions of $SU(N)$ Vafa-Witten theory are determined
for arbitrary $N$ in \cite{Manschot:2014cca}. These expressions give
rise to higher dimensional analogues of Appell functions.  
For $SU(3)$, the complete non-holomorphic partition function has a
similar form to (\ref{pfsu2}), it reads \cite{Manschot:2010nc,
  Manschot:2011ym, Manschot:2017xcr}:
\begin{eqnarray}
Z^{\mathbb{CP}^2}_{3,\mu}(\tau,\bar \tau) &=& \frac{\widehat f_{3,\mu}(\tau,\bar\tau)}{\eta(\tau)^{9}},
\end{eqnarray}
where $\widehat f_{3,\mu}$ reads
\be
\label{hatf3mu}
\begin{split}
\widehat f_{3,\mu}(\tau,\bar\tau)&=f_{3,\mu}(\tau)-\frac{i \left(3/2 \right)^{3/2}}{\pi}
\sum_{\nu=0,1} \int_{-\bar \tau}^{i\infty}
\frac{\widehat
  f_{2,\nu}(\tau,-v)\,\Theta_{\frac{\mu}{3}+\frac{\nu}{2}}(3v)}{(-i(v+\tau))^{3/2}}\,dv,
\end{split}
\ee
with $\widehat f_{2,\nu}$ as in (\ref{hatf2mu}). The first few terms
of the $q$-expansions of $f_{3,\mu}(\tau)$ are:
\be
\begin{split}
f_{3,0}(\tau) &= \frac{1}{9}-q+3q^2 +\dots, \\
f_{3,1}(\tau) &=f_{3,2}(\tau)= 3q^{5/3}+15q^{8/3}+36q^{11/3}+\dots.
\end{split}
\ee
See Table \ref{coef} in Section \ref{sec5} for a longer list. 
The holomorphic parts of $\widehat f_{3,\mu}$ transform as:
\be 
\begin{split}
  f_{3,\mu}(-1/\tau) &= \frac{i\tau^{3}}{\sqrt{3}}\sum_{\nu=0}^2
                      e^{-2\pi i\mu\nu/3}\\
  &\times
  \left(f_{3,\nu}(\tau)-\frac{i \left(3/2 \right)^{3/2}}{\pi}\sum_{\alpha=0,1}\int_{0}^{i\infty}\frac{\widehat
      f_{2,\alpha}(\tau,-w)\,\Theta_{\frac{2\nu+3\alpha}{6}}(3w)}{(-i(w+\tau))^{3/2}}\;dw\right),\\
f_{3,\mu}(\tau+1)&=(-1)^\mu e^{\pi i \mu^2/3}f_{3,\mu}(\tau).
  \label{f3mu}
\end{split}
\ee
The non-holomorphic parts satisfy a compact holomorphic anomaly
equation \cite{Minahan:1998vr, Alexandrov:2016tnf, Alexandrov:2019rth,
  Bonelli:2020xps, Dabholkar:2020fde, Manschot:2021qqe}.

In the next section, we will introduce various notions of modular forms and
mock modular forms to characterize the various functions, which
appeared above.

\section{Modular forms and mock modular forms}\label{sec3}
In this section, we introduce the definitions of modular and mock
modular forms for $SL_2(\mathbb{Z})$ and their vector-valued
counterparts. We also introduce the notion of mock cusp forms.
\subsection{Modular, mock modular and mock cusp forms}\label{sec31}
We start with the basic definition of a modular form:\\
\\
\noindent 
{\bf Definition:} 
{\it A modular form of weight $k$ is a holomorphic function $f(\tau):\mathbb{H}\to
  \mathbb{C}$, which
\begin{enumerate}
\item   transforms under an $SL_2(\mathbb{Z})$ matrix $\begin{pmatrix}
a & b\\ c & d \end{pmatrix}$ as follows:
\begin{eqnarray}
f\left(\frac{a\tau+b}{c\tau+d}\right)=(c\tau+d)^kf(\tau),
\end{eqnarray}
\item and whose growth for $\tau\to i\infty$ is such that
  \be
  \label{cuspgrowth}
\lim_{\tau\to i\infty} (c\tau+d)^{-k} f\left(\frac{a\tau+b}{c\tau+d}\right)
\ee
is bounded for all $\begin{pmatrix}
a & b\\ c & d \end{pmatrix}\in SL_2(\mathbb{Z})$.
\end{enumerate}
}
 There are two generators of $SL_2(\mathbb{Z})$, namely $T=\begin{pmatrix}
  1 & 1\\ 0 & 1
 \end{pmatrix}$ and $S=\begin{pmatrix}
 0 &-1\\1 & 0
 \end{pmatrix}$. Under these transformations
 \begin{eqnarray}
 f(\tau+1) =f(\tau),\quad f\left(-\frac{1}{\tau}\right)=\tau^k f(\tau).
 \end{eqnarray}
 Due to the symmetry under the $T$-transformation, $f(\tau)$ can be expanded as a Fourier
 series. This series starts with a constant term $a_0 $ as a result of
 the growth condition, $$f(\tau)=\sum_{n=0}^{\infty}a(n)\, q^n,\qquad q=e^{2\pi
   i\tau}.$$
 One can allow for phases in the transformations 
 \be 
\begin{split}
& f(\tau+1) = \varepsilon(T)\, f(\tau),\\ \nn
& f\left(\frac{-1}{\tau}\right) = \varepsilon(S)\, \tau^k f(\tau).
  \end{split}
 \ee
Since $S^2=(ST)^3=-I$ we have $\varepsilon(S)=\varepsilon(T)^{-3}$.
 \\
\\
We next introduce the notion of a cusp form:
\\
\\
\noindent 
{\bf Definition:} 
{\it A modular form $f(\tau)$ for $SL_2(\mathbb{Z})$ is a cusp form
  for which the combination (\ref{cuspgrowth}) vanishes for $\tau\to
  i\infty$. As a result, the constant term $a_0$ of its Fourier series vanishes.
}
\\
\\
The space of modular forms is well-known to be finite dimensional for
fixed weight $k$. The Petterson innerproduct
\be
\label{PettersonInn}
\left<f,g\right>=\int_{\mathbb{H}/SL_2(\mathbb{Z})} \frac{d\tau\wedge
  d\bar \tau}{y^{2-k}} f(\tau)\,\overline{g(\tau)},
\ee
forms a natural innerproduct on the space of cusp forms. Moreover,
$\left<f,g\right>$ vanishes for $f$ an Eisenstein series and $g$ a
cusp form.

The definitions of a modular form and cusp are readily extended from
the full modular group $SL_2(\mathbb{Z})$ to a congruence subgroup $\Gamma\subset
SL_2(\mathbb{Z})$.
Another useful notion is a vector-valued modular form:
\noindent
\vspace{.3cm}\\
{\bf Definition:} {\it A $d$-dimensional vector-valued modular form of
  weight $k$ under $SL_2(\mathbb{Z})$ is a vector of holomorphic functions
\be
 \vec f=\left[ \begin{array}{c} f_0 \\ \vdots \\ f_{d-1}\end{array}\right]:\mathbb{H}\to \mathbb{C}^d,
\ee
 with the following  properties:
\begin{enumerate}
\item The elements of the vector transform under the $S$ and $T$
  transformations as
\begin{eqnarray}
S:\qquad \vec f\left(\frac{-1}{\tau}\right) &=&  \tau^k\, {\bf M}(S)
                                                \vec f(\tau),\\
T:\qquad \vec f(\tau+1) &=& {\bf M}(T)\vec f(\tau).
\end{eqnarray}
with ${\bf M}(S)$ and ${\bf M}(T)$ as $d\times d$ matrices. ${\bf M}(T)$
is diagonal in many cases. 
\item For each element $f_\mu$, $\mu=0,\dots,d-1$, the combination
  (\ref{cuspgrowth}) is bounded for all matrices
  $\left(\begin{array}{cc} a & b \\ c & d  \end{array}\right)\in SL_2(\mathbb{Z})$. 
\end{enumerate}}
\noindent This notion is readily extended to a vector-valued cusp
form. The elements $f_\mu$ of the vector $\vec f$ are modular forms
for a congruence subgroup $\Gamma\subset SL_2(\mathbb{Z})$.
\\
\\
\noindent We next continue with the definition of mock modular form
\cite{zagier:75,ZwegersThesis,zagier2}. We first introduce a map on
non-holomorphic functions. For a function $g:\mathbb{H}\times 
\mathbb{\bar H}\to \mathbb{C}$, we define the function $g^*:\mathbb{H}\times 
\mathbb{\bar H}\to \mathbb{C}$ by:
\begin{eqnarray}
  \label{gstar}
g^*(\tau,\bar{\tau})=-2^{1-k}i\int_{-\bar{\tau}}^{i\infty} \frac{g(\tau,-v)}{(-i(v+\tau))^{k-\ell}}\, dv\;.
\end{eqnarray}
Assuming that $g$ transforms under an element of $\begin{pmatrix}
a & b\\ c & d
\end{pmatrix}\in SL_2(\mathbb{Z})$ as
\be
g\left(\frac{a\tau+b}{c\tau+d},\frac{au+b}{cu+d}\right) = (c\tau+d)^k (cu+d)^{2-k+\ell} g(\tau,u),
\ee
for some $k$ and $\ell$, the function $g^*$ transforms as,
\be
g^*\left(\frac{a\tau+b}{c\tau+d},\frac{a\bar{\tau}+b}{c\bar{\tau}+d}\right) =(c\tau+d)^k\left( g^*(\tau,\bar{\tau})+2^{1-k}i\int_{d/c}^{i\infty} \frac{g(\tau,-v)}{(-i(v+\tau))^{k-\ell}}\, dv \right).
\ee
\vspace{.3cm}\\
{\bf Definition:}
{\it
A mock modular form of weight $k$ is a holomorphic $q$-series $f:\mathbb{H}\to
\mathbb{C}$, such that its completion
\be
\label{defmock}
\widehat f(\tau,\bar \tau)=f(\tau)+g^*(\tau,\bar \tau)
\ee
\begin{enumerate}
\item transforms as a modular form of weight $k$,
\item $g^*$ is the image
under the map (\ref{gstar}) of the complex conjugate of a modular form 
with weight $2-k$,
\item The combination (\ref{cuspgrowth}) is bounded for $f$ and all matrices
  $\left(\begin{array}{cc} a & b \\ c & d  \end{array}\right)\in SL_2(\mathbb{Z})$. 
\end{enumerate}
}

There are many variations to the above definitions. A mixed mock modular form if a function
$f$ as above but with $g$ a product of a modular form of weight $k$
and the complex conjugate of a modular form of weight $2-k+\ell$ for
some $k$ and $\ell$. Similarly, one can consider sums of products.

Clearly, the function $g$ is crucial information to characterize the
mock modular form $f$. It is called  the shadow of $f$, and can be obtained by taking a non-holomorphic derivative of $\widehat f$,
\be
\label{defshadow}
g=y^{k-\ell}\partial_{\bar{\tau}}\widehat f,
\ee
with $y={\rm Im}(\tau)$. Thus the
shadow is an element of
$\overline{{M}}_{2-k+\ell}(\Gamma)$ for some group $\Gamma\subset SL_2(\mathbb{Z})$, that is to say the shadow is the complex
conjugate of a modular form of weight $2-k+\ell$.
 
A further interesting extension of mock modular forms are those of
depth $r\geq 1$. Mock modular forms of depth 1 are the functions defined
in the previous definition and equation (\ref{defmock}). The notion of mock modular forms of depth $r$ is defined
  iteratively for $r\geq 1$ \cite{Nazaroglu:2016lmr, Larry,
    Manschot:2017xcr}.  To explain the definition, we introduce the vector spaces $\mathbb{M}_k ^r(\Gamma)$ of
mock modular forms of depth $r\geq 1$, with weight $k$ for the group $\Gamma\subset
SL_2(\mathbb{Z})$. We furthermore let
$\mathbb{M}_k ^{-1}(\Gamma)=\emptyset$, and $\mathbb{M}_k
^0(\Gamma)=M_k(\Gamma)$, i.e. the space of classical modular forms of
weight $k$. We furthermore introduce the vector spaces
$\widehat{\mathbb{M}}_k ^r$ of the completed, non-holomorphic functions of
$\mathbb{M}_k ^r(\Gamma)$. We then have the following definition:
\noindent
\vspace{.3cm}\\
{\bf Definition:}
{\it A mock modular form of depth $r$ and with weight $k$ for the
  group $\Gamma\subset SL_2(\mathbb{Z})$ is defined by the property
  that its shadow is an element of $\widehat{\mathbb{M}}_k ^{r-1}\otimes \overline{{M}}_{2-k+\ell}$.
  }
\vspace{.2cm}
  \\
We introduce moreover the notion of a mock modular cusp form.
\noindent
\vspace{.3cm}\\
{\bf Definition:} {\it 
  A mock cusp form of depth $r$ is a mock modular form $f(\tau)$ of
  depth $r$ such that the
  combination (\ref{cuspgrowth}) vanishes for all elements of
  $SL_2(\mathbb{Z})$. Thus in particular the constant term of the
  Fourier series of $f$ vanishes.
}
\\
\\
The notion of mock modular form and mock cusp form are readily
extended to vector-valued mock modular forms and vector-valued mock
cusp forms.

\subsection{VW partition functions and (mock) modular
  forms}\label{sec32}

With the terminology developed in above, we can characterise the
functions appearing Section \ref{sec22}. Clearly, the Dedeking eta
function (\ref{Deta}) is a modular form of weight 1/2 with $\varepsilon(T)=e^{2\pi
  i/24}$. Furthermore, $\Theta_\mu$, $\mu=0,1$, are modular forms of weight 1/2
for the congruence subgroup $\Gamma_0(4)$. The vector $\vec \Theta$ of
the two functions,
\be
\vec \Theta(\tau)=\left[\begin{array}{c} \Theta_0(\tau) \\ \Theta_1(\tau) \end{array}\right]
\ee
is a vector-valued modular form  for $SL_2(\mathbb{Z})$.

The vector $\vec f_{2}$ of the functions $f_{2,\mu}$ (\ref{f2mu}),
\be
\vec f_2(\tau)=\left[\begin{array}{c} f_{2,0}(\tau) \\ f_{2,1}(\tau) \end{array}\right],
\ee
is a vector-valued mock modular form of weight 3/2 and depth 1. Moreover, the vector
$\vec f_{3}$ of the functions $f_{3,\mu}$ (\ref{f2mu}) is a
vector-valued mock modular form of weight 3 and depth 2 \cite{Manschot:2017xcr}.

To further study these functions, we first introduce the theta series
\be
b_{3,j}(\tau) = \sum_{k_1,k_2\in\mathbb{Z}+j/3}q^{k_1^2+k_2^2+k_1k_2}.
\ee
The modular transformations of these functions are:
\begin{eqnarray}
b_{3,j}(-\frac{1}{\tau}) &=& -\frac{i\tau}{\sqrt{3}}\sum_{\ell=0}^2 e^{-2\pi ij\ell/3}b_{3,\ell}(\tau), \\ \nn
b_{3,j}(\tau+1) &=& e^{2\pi ij^2/3}b_{3,j}(\tau).
\end{eqnarray}
We then form the 3-dimensional vector-valued modular form $\vec m: \mathbb{H}\to \mathbb{C}^3$,
\begin{eqnarray}
\vec m=\begin{bmatrix}
m_0\\m_1\\m_2
\end{bmatrix}
=\begin{bmatrix}
\frac{b_{3,0}^3+2b_{3,1}^3}{9}\\
\frac{b_{3,0}b_{3,1}^2}{3}\\
\frac{b_{3,0}b_{3,1}^2}{3}\end{bmatrix},
\end{eqnarray}
The transformations of $m_{\mu}$ under the generators of $SL_2(\mathbb{Z})$ are given by,
\be
\begin{split}
  m_{\mu}(\tau+1) &= (-1)^{\mu}e^{\pi i\mu^2/3}m_{\mu}(\tau),\\ 
m_{\mu}\!\left(-\frac{1}{\tau}\right) &= \frac{i\tau^3}{\sqrt{3}}\sum_{\nu=0}^2 e^{-2\pi i\nu\mu/3}m_{\nu}(\tau).
\end{split}
\ee
These transformations are identical to those of the completed mock
modular form $\widehat f_{3,\mu}$ (\ref{hatf3mu}) obtained from
$SU(3)$ Vafa-Witten theory. There is no analog of this type of purely
holomorphic functions for $SU(2)$, since a holomorphic modular form
with the same transformations as $\widehat f_{2,\mu}$ does not exist \cite{Manschot:2008zb}.

The first few terms in the series expansion of $m_{\mu}$ can be given by,
\begin{eqnarray}
m_0(\tau) &=& \frac{1}{9}+8q+30q^2 +\dots\\ \nn
m_1(\tau) &=& 3q^{2/3}+24q^{5/3}+51q^{8/3}+\dots
\end{eqnarray}
We can in fact explicitly write the series for $m_0(\tau)$, and $m_1(\tau)$ as follows:
\be
\begin{split} 
\label{mmucoeff}
  m_0(\tau) &= \frac{1}{9}+\sum_{n=0\;{\rm mod}\;3\atop n>0}\sum_{d|n}\chi_{n,d} \;d^2 q^{n/3},\\ 
m_1(\tau) &= \sum_{n=2\;{\rm mod}\;3}\sum_{\begin{smallmatrix}
\frac{n}{d}=1\;{\rm mod}\;3\\
d\in\mathbb{Z}
\end{smallmatrix}} {\rm sgn}(d)\; d^2 q^{n/3},
\end{split}
\ee
where, the character $\chi_{n,d}$ is given for $n>0$ by,
\begin{equation}
\chi_{n,d}=\begin{cases}
      -1 & \text{if $d=1$ mod 3, $n/d=0$ mod $3$}\\
      1 & \text{if $d=2$ mod 3, $n/d=0$ mod $3$}\\
      1 & \text{if $d=0$ mod 3, $n/d=1$ mod $3$}\\
      -1 & \text{if $d=0$ mod 3, $n/d=2$ mod $3$}\\
      0 & \text{otherwise}.
    \end{cases}  
\end{equation}
As a result, the coefficient of $q^{p/3}$ of $m_1(\tau)$ with $p$ a
prime is given by $p^2-1$. The coefficients $d_0(p)$ of $q^p$ of $m_0(\tau)$ with $p$ prime are of two types,
\begin{enumerate}
\item For $p=2 \;{\rm mod}\; 3$,
  \be
  \label{d0p2}
d_0(p)=10p^2-10,
\ee
\item For $p=1 \;{\rm mod}\; 3$,
  \be
  \label{d0p1}
d_0(p)=8p^2+8.
\ee
For example $d_0(3)=80$.
\end{enumerate}
We observe that coefficients of $q^p$ for prime $p$ in $m_0$ are bounded by $10p^2$. 

Now having described $\vec m$, we can obtain an example $\vec S$ of a
vector-valued mock cusp
form of depth 2. Namely, we define $\vec S$ as
\be
\label{defvecS}
\vec S=\frac{1}{3}(\vec m-\vec f).
\ee
The first terms in the $q$-series of
$S_0$ and $S_1$ are:
\begin{eqnarray}
\label{Smucoeffs}
S_0(\tau) &=& 3q+9q^2+21q^3+\dots,\\ \nn
S_1(\tau) &=& S_2(\tau)=q^{2/3}+7q^{5/3}+\dots.
\end{eqnarray}
The constant terms of $S_\mu$ thus vanish as for the classical cusp
forms. More terms are listed in Table \ref{coef} in Section
\ref{sec5}. 
 
Moreover with Eq. (\ref{f3mu}), we see that the elements $S_\mu$ of $\vec S$ transform as
\be
\begin{split}
  S_{\mu}(-1/\tau) &= \frac{i\tau^{3}}{\sqrt{3}}\sum_{\nu=0}^2
                      e^{-2\pi i\mu\nu/3}\\
  &\times
  \left(S_{\nu}(\tau)+\frac{\sqrt{3}i}{2\sqrt{2}\pi}\sum_{\alpha=0,1}\int_{0}^{i\infty}\frac{\widehat
      f_{2,\alpha}(\tau,-w)\,\Theta_{\frac{2\nu+3\alpha}{6}}(3w)}{(-i(w+\tau)) ^{3/2}}\;dw\right),\\
S_{\mu}(\tau+1)&=(-1)^\mu e^{\pi i \mu^2/3}S_{\mu}(\tau).
  \label{S3mu}
\end{split}
\ee
Since 
\be
\lim_{\tau\to \infty} \int_{0}^{i\infty}\frac{\widehat
      f_{2,\alpha}(\tau,w)\,\Theta_{\frac{2\nu+3\alpha}{6}}(3w)}{(-i(w+\tau)^{3/2})}\;dw=0,
\ee
the combination (\ref{cuspgrowth}) vanishes for
$S_{\mu}$. We thus confirm that $S_\mu$ are mock cusp forms for the congruence
subgroup $\Gamma(3)\subset SL_2(\mathbb{Z})$. It would be interesting
to understand better the spaces of such functions, and for example
define a suitable inner product. The standard innerproduct
(\ref{PettersonInn}) diverges for functions such as $\widehat
S_\mu(\tau,\bar \tau)$, and a
suitable regularization will need to be defined. We leave this for
future work.

We have some interesting observations for the growth of the Fourier coefficients of these functions. These are shown as plots in the next section. 
In section \ref{sec4} we briefly discuss the behavior of the growth of coefficients of cusp forms and how they might change when there is a non-holomorphic piece in the modular transformation.

\section{Numerical experiments for the coefficients of VW partition functions}\label{sec5}
This section carries out varies numerical experiments on the
coefficients of $f_{2,\mu}$, $f_{3,\mu}$, $m_\mu$ and
$S_\mu$. Especially for $S_{\mu}$ we find some intriguing pattern,
namely the prime coefficients appear to be well approximated by a
constant times $p^{3/2}$. 

\subsection{Coefficients for $N=2$}
We study in this subsection the coefficients of
$f_{2,\mu}=3G_{2,\mu}$. Since the coefficients of $G_{2,\mu}$ are the well-known
Hurwitz class numbers, we will focus on these. We define for $\mu=0,1$:
\be
G_{2,\mu}(\tau) = \sum_{n\in\mathbb{Z}-\frac{\mu}{4}} a_{\mu}(n) q^{n}.
\ee
We first tabulate the coefficients for of $G_{2,\mu}$ up to $n=45$ in Table
\ref{coefclass}. 
\newpage 
\begin{longtable}{|p{.20\textwidth}|p{.20\textwidth}|p{.20\textwidth} | }
\hline 
$n$ & $a_0(n)$ & $a_1(n-1/4)$ \\
\hline\hline
 0 & $-\frac{1}{12}$ & 0 \\
 1 & $\frac{1}{2}$ & $\frac{1}{3}$ \\
 2 & 1 & 1 \\
 3 & $\frac{4}{3}$ & 1 \\
 4 & $\frac{3}{2}$ & 2 \\
 5 & 2 & 1 \\
 6 & 2 & 3 \\
 7 & 2 & $\frac{4}{3}$ \\
 8 & 3 & 3 \\
 9 & $\frac{5}{2}$ & 2 \\
 10 & 2 & 4 \\
 11 & 4 & 1 \\
 12 & $\frac{10}{3}$ & 5 \\
 13 & 2 & 2 \\
 14 & 4 & 4 \\
 15 & 4 & 3 \\
 16 & $\frac{7}{2}$ & 5 \\
 17 & 4 & 1 \\
 18 & 3 & 7 \\
 19 & 4 & $\frac{7}{3}$ \\
 20 & 6 & 5 \\
 21 & 4 & 3 \\
 22 & 2 & 6 \\
 23 & 6 & 2 \\
 24 & 6 & 8 \\
 25 & $\frac{5}{2}$ & 3 \\
 26 & 6 & 5 \\
 27 & $\frac{16}{3}$ & 3 \\
 28 & 4 & 8 \\
 29 & 6 & 2 \\
 30 & 4 & 10 \\
 31 & 6 & 2 \\
 32 & 7 & 5 \\
 33 & 4 & 5 \\
 34 & 4 & 8 \\
 35 & 8 & 3 \\
 36 & $\frac{15}{2}$ & 10 \\
 37 & 2 & $\frac{7}{3}$ \\
 38 & 6 & 7 \\
 39 & 8 & 4 \\
 40 & 6 & 10 \\
 41 & 8 & 1 \\
 42 & 4 & 11 \\
 43 & 4 & 5 \\
 44 & 10 & 7 \\
 45 & 6 & 5 \\
 \hline
	\caption{First few coefficients $a_\mu(n)=H(4n-\mu)$ of the class number generating functions $G_\mu$ \eqref{f2021series}.}\label{coefclass}
\end{longtable}

\begin{figure}[H]
 \includegraphics{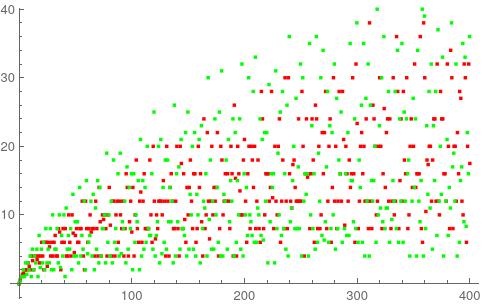}
 \caption{Plot of the coefficients $a_\mu(n)=H(4n-\mu)$ of $G_\mu$ as
   function of $n$. The red dots represent the coefficients $a_0(n)$, while the green ones represent the coefficients $a_1(n-1/4)$.}
\label{ClassNrPlot}
\end{figure}

We observe from the table that the behavior of the coefficients as function of $n$
is not monotonic, but that the coefficients grow on average. To get a better impression, we plot the coefficients up to
$n=400$ in Figure \ref{ClassNrPlot}. We observe that the coefficients
are highly scattered, but on average appear to grow as a power law. 

\begin{figure}[H]
 \includegraphics{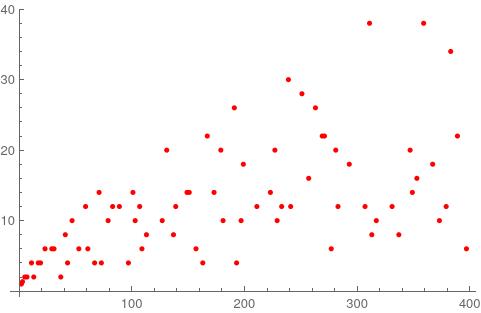}
 \caption{Diagram with the coefficients $a_0(p)$ as function of prime
   $p$.}
 \label{aoprime}
\end{figure}

\begin{figure}[H]
 \includegraphics{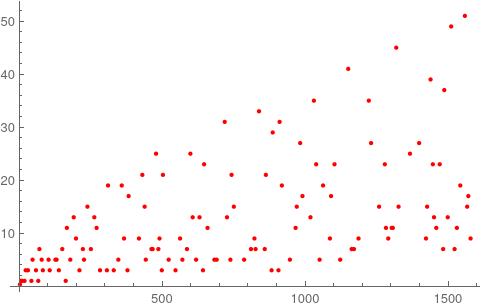}
 \caption{Diagram with the coefficients $a_1(p/4)$ as function for
   prime $p$ of type $4n+3$.}
 \label{a1prime}
\end{figure}

Now in many cases, such as Eisenstein series of integer weight, the
growth of the coefficients is more regular for prime numbers. This we
for example encountered for the functions $m_\mu$ (\ref{mmucoeff}). We
therefore present plots for $a_\mu(p)$ as function of prime $p$ in
Figures \ref{aoprime} and \ref{a1prime}.
   
\subsection{Coefficients for $N=3$}
We proceed in this subsection with partition functions for $SU(3)$,
and compare our findings with those for $N=2$. We define the
coefficients $c_\mu(n)$, $d_\mu(n)$ and $s_\mu(n)$ through
\begin{eqnarray}\label{defs}
f_{3,\mu}(\tau) &=& \sum_{n\in\mathbb{Z}-\frac{\mu}{3}} c_{\mu}(n) q^{n},\\ \nn
m_{\mu}(\tau) &=& \sum_{n\in\mathbb{Z}-\frac{\mu}{3}} d_{\mu}(n) q^{n},\\ \nn
S_{\mu}(\tau) &=& \sum_{n\in\mathbb{Z}-\frac{\mu}{3}} s_{\mu}(n) q^{n}.
\end{eqnarray}

First we list the
coefficients of $f_{3,\mu}$ and $S_{\mu}$ in Table \ref{coef}. Since
$S_\mu$ is a mock cusp form, we expect that the coefficients of are
smaller than those for $f_\mu$ for $n\gg 1$. We observe from the table
that this is indeed the case.


\begin{longtable}{|p{.15\textwidth}|p{.15\textwidth}|p{.15\textwidth}|p{.15\textwidth} |p{.15\textwidth} | }
\hline 
$n$ & $c_0(n)$ & $c_1(n-\frac{1}{3})$ & $s_0(n)$ & $s_1(n-\frac{1}{3})$\\
\hline\hline
 0 & $\frac{1}{9}$ & 0 & 0 & 0 \\
 1 & $-1$ & 0 & 3 & 1 \\
 2 & 3 & 3 & 9 & 7 \\
 3 & 17 & 15 & 21 & 12 \\
 4 & 41 & 36 & 21 & 28 \\
 5 & 78 & 69 & 54 & 27 \\
 6 & 120 & 114 & 42 & 58 \\
 7 & 193 & 165 & 69 & 49 \\
 8 & 240 & 246 & 90 & 94 \\
 9 & 359 & 303 & 123 & 69 \\
 10 & 414 & 432 & 54 & 136 \\
 11 & 579 & 492 & 207 & 109 \\
 12 & 626 & 669 & 138 & 177 \\
 13 & 856 & 726 & 168 & 120 \\
 14 & 906 & 975 & 198 & 235 \\
 15 & 1194 & 999 & 258 & 187 \\
 16 & 1172 & 1332 & 156 & 292 \\
 17 & 1638 & 1338 & 414 & 155 \\
 18 & 1569 & 1743 & 207 & 355 \\
 19 & 1987 & 1716 & 303 & 278 \\
 20 & 2040 & 2226 & 360 & 418 \\
 21 & 2578 & 2130 & 474 & 252 \\
 22 & 2340 & 2775 & 180 & 435 \\
 23 & 3255 & 2625 & 675 & 373 \\
 24 & 2940 & 3354 & 414 & 562 \\
 25 & 3665 & 3129 & 381 & 327 \\
 26 & 3642 & 4041 & 486 & 653 \\
 27 & 4490 & 3735 & 690 & 395 \\
 28 & 3940 & 4752 & 420 & 712 \\
 29 & 5484 & 4317 & 972 & 411 \\
 30 & 4734 & 5532 & 342 & 796 \\
 31 & 5815 & 5070 & 627 & 598 \\
 32 & 5814 & 6393 & 792 & 765 \\
 33 & 7014 & 5694 & 942 & 553 \\
 34 & 5832 & 7317 & 360 & 961 \\
 35 & 8274 & 6582 & 1242 & 696 \\
 36 & 7115 & 8277 & 783 & 1057 \\
 37 & 8566 & 7272 & 798 & 456 \\
 38 & 8322 & 9345 & 846 & 1141 \\
 39 & 10018 & 8325 & 1194 & 865 \\
 40 & 8334 & 10425 & 486 & 1325 \\
 41 & 11778 & 9087 & 1674 & 693 \\
 42 & 9708 & 11541 & 864 & 1161 \\
 43 & 11785 & 10281 & 1005 & 942 \\
 44 & 11604 & 12855 & 1332 & 1435 \\
 45 & 13614 & 11058 & 1302 & 804 \\
 46 & 10998 & 14175 & 558 & 1531 \\
 47 & 15843 & 12327 & 2079 & 1091 \\
 48 & 13178 & 15486 & 1074 & 1638 \\
 49 & 15531 & 13263 & 1359 & 909 \\
 50 & 14817 & 16959 & 1071 & 1747 \\
 \hline
	\caption{First few coefficients of the mock modular forms $f_{3,\mu}$ and the mock cusp forms $S_{\mu}$}\label{coef}
\end{longtable}

\begin{figure}[H]
 \includegraphics{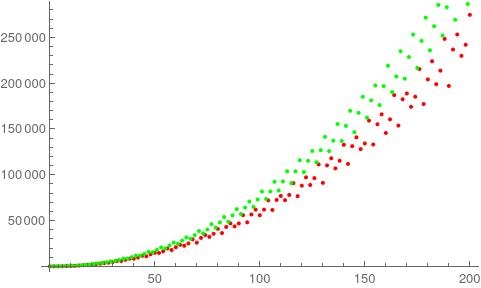}
 \caption{Green dots represent the coefficients $c_0(n)$ for odd $n$
   while the red ones represent the coefficients $c_0(n)$ for even
   $n$.}
 \label{coeffsf0}
\end{figure}

\begin{figure}[H]
 \includegraphics{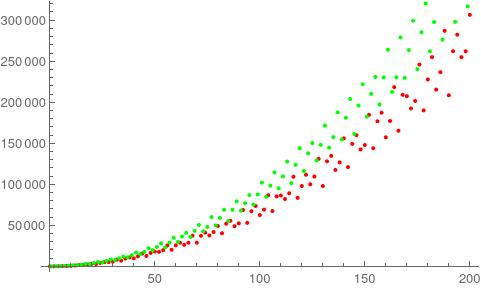}
 \caption{Green dots represent the coefficients $d_0(n)$ for odd $n$
   while the red ones represent the coefficients $d_0(n)$ for even
   $n$.}
 \label{coeffsm0}
\end{figure}

To get a better impression of the growth, we plot the coefficients of
the functions for a larger range, starting with $f_0$ in Figure \ref{coeffsf0}. We observe from
the table that the magnitude of the coefficients alternate between 
even and odd $n$. We therefore distinguish the even and odd
coefficients with red and green respectively. We observe from the plot
in Figure \ref{coeffsf0} that the even and odd coefficients remain
separate. This behavior is not unique to $f_0$. Figure \ref{coeffsm0}
demonstrates a similar behavior for the coefficients of $m_0$.

We have seen in Section \ref{sec32} that the prime coefficients of
$m_\mu$ have a very simple expression. See Eqs (\ref{d0p2}) and
(\ref{d0p1}). This is a general property of Eisenstein series. Let us
therefore plot the prime coefficients $d_0(p)$ for $m_0$, and $c_0(p)$
of $f_{3,0}$. We observe that the plot is far less scattered than the
original plots for both $m_0$ and $f_{3,0}$. We also observe that the
growth in this range is roughly comparable. We will derive an
upperbound for the growth in the next section, as well as discuss an
average. 

\begin{figure}[H]
 \includegraphics{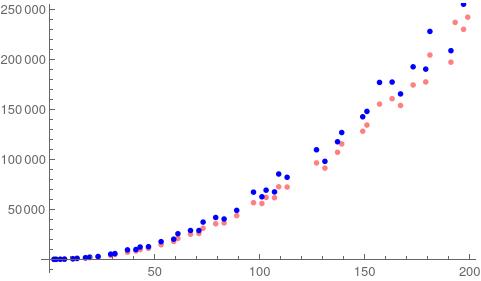}
 \caption{Blue dots represent the coefficients $d_0(p)$ of
   $m_{0}(\tau)$, and the pink ones represent the coefficients $c_0(p)$ of $f_{3,0}(\tau)$ for prime $p$.}
\end{figure}

\begin{figure}[H]
 \includegraphics{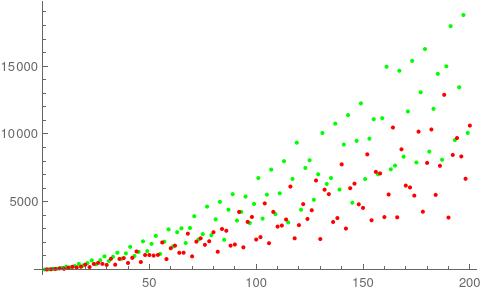}
 \caption{Green dots represent the coefficients $s_0(n)$ for odd $n$
   while the red ones represent the coefficients $s_0(n)$ for even
   $n$.}
 \label{coeffsS0}
\end{figure}

Next we consider the coefficients of the mock cusp form,
$S_0(\tau)=\frac{1}{3}(m_{0}-f_{3,0})$, which are possibly of most
interest. Its coefficients $s_\mu(n)$ are plotted
in Figure \ref{coeffsS0}, again distinguishing even and odd $n$. We
observe that coefficients are very scattered, and that the two sets of
coefficients are not separated as was the case for $f_{3,0}$ or
$m_0$. Still, the coefficients of odd powers of $q$ are typically
larger than those of even powers of $q$.

\begin{figure}[H]
 \includegraphics{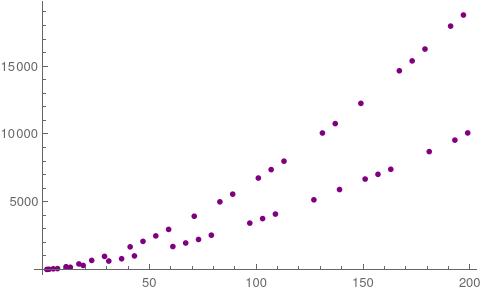}
 \caption{The dots $s_0(p)$ for prime $p$. To note that from 11 upto
   199 all twin primes show the following behavior, if $p$ and $p+2$
   are twin primes then $s_0(p)>s_0(p+2)$.}
 \label{s0prime}
\end{figure}

As mentioned above, the behavior of the coefficients is in general
better for the prime coefficients. We plot the coefficients
$s_0(p)$ for prime $p$ in Figure \ref{s0prime}. Here we do see a
striking behavior, namely that these points appear to lie on a smooth
curve. Such a regular curve is not generally the
case for classical cusp forms, such as $\eta^{24}$.

\begin{figure}[H]
 \includegraphics{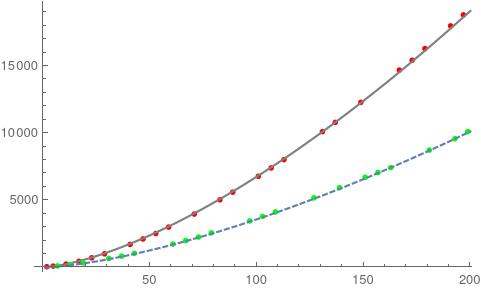}
 \caption{Least square fits for $s_0(p)$ with primes $p$ of the form $3n\pm 1$. The red dots represent the coefficient $s_0(p)$ for prime $p=3n-1$ and the grey line is given by $6.75467\,x^{3/2}$, the green dots represent $s_0(p)$ for prime $p=3n+1$ and the dashed line is given by $3.57843\,x^{3/2}$.}
\label{soprimelsq}
\end{figure}

We make a curve fit in Figure \ref{soprimelsq}. The least square fit
in the last plot (except for $p=3$), suggests that the coefficients
grow as $n\sim n^{3/2}$. Naturally, it is desirable to prove this
growth. In Section \ref{sec4}, we give an upperbound, which is
unfortunately much weaker.
  
\begin{figure}[H]
 \includegraphics{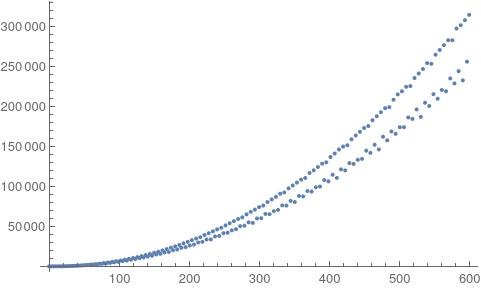}
 \caption{Plot of the coefficients $c_1(n+2/3)$ of $f_{3,1}$. The
   numbers on the horizontal axis are $3n+2$ for integer $n\leq 200$.}
\label{coefff31}
\end{figure}

Next we repeat the above plots for the functions with $\mu=1$. Figure
\ref{coefff31} plots the coefficient $c_1(n+2/3)$ of $f_{3,1}$. We again
observe different bands as in Figure \ref{coeffsf0} for $f_{3,0}$. The
plot here appears less scattered though.

\begin{figure}[H]
 \includegraphics{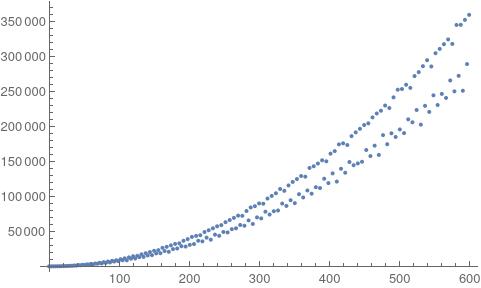}
 \caption{Plot of the coefficients $d_1(n+2/3)$ of $m_{1}$. The
   numbers on the horizontal axis are $3n+2$ for $n\leq 200$.}
\label{coeffsm1}
\end{figure}

We continue with the plot for the coefficients $d_1(n+2/3)$ of $m_1$
in Figure \ref{coeffsm1}. The plot is quite similar to Figures
\ref{coeffsm0} and \ref{coefff31}. The curves are quadratic since for
$3n+2$ equal to a prime $p$, $d_1(p/3)=p^2-1$. The prime coefficients
of $c_1(p/3)$ of $f_{3,1}$ are also well approximated by a parabolic curve.

\begin{figure}[H]
 \includegraphics{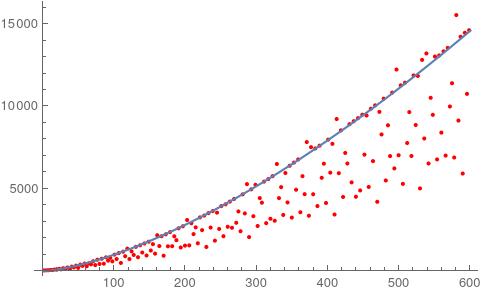}
 \caption{Plot of coefficients $s_1(n+2/3)$ with the numbers $3n+2$ on
   the horizontal axis. The least square fit for the prime
   coefficients $s_1(p/3)$ is given by $0.993519\, x^{3/2}$.}
\label{coeffsS1}
\end{figure}

Lastly, we consider the coefficients $s_1(n+2/3)$ of the mock cusp
form $S_1$. We plot in Figure \ref{coeffsS1} the coefficients
$s_1(n+2/3)$, and the least square fit for the prime coefficients
$s_1(p/3)$. Similar to the $S_0$, the least square fit suggests that
the coefficients grow as $\sim n^{3/2}$. This is notably larger than
Deligne's bound $\sim n^1$ for the coefficients of proper cusp forms of weight
3. Thus the modified transformation of $S_\mu$ involving the period integral
must have an important impact on the growth of the coefficients.

\section{Asymptotics of Fourier coefficients}\label{sec4}
This section reviews the calculation of upper bounds on the growth of Fourier
coefficients of modular forms and cusp forms using the saddle point
method. We will then generalize this analysis to mock modular forms
and cusp forms, in particular to the function $f_{2,\mu}$, $f_{3,\mu}$
and $S_\mu$. 

It is well known that the saddle point method (or
Hecke bound) typically gives a very large upperbound, that is to say
order $n^k$ for
modular forms, whereas the correct bound is order $n^{k-1}$. For cusp
forms the saddle point method gives $n^{k/2}$. On the other hand, the sharper Ramanujan-Petersson conjecture, proved by Deligne in 
\cite{deligne:74,deligne:80}, states that the Fourier coefficients $a(n)$ of a holomorphic cusp forms for $SL_2(\mathbb{Z})$ bounded by
$$|a(n)| < \sigma_0(n)n^{\frac{k-1}{2}},$$ where $\sigma_0(n)$ is the number of divisors of $n$. Thus for prime numbers the bound is given by,
$$|a(p)| < 2p^{\frac{k-1}{2}},$$ where $k$ is the weight of the cusp form. 

Similarly,
we do find that the saddle point analysis for mock modular and mock
cusp forms gives rise to larger exponents than
those observed in the previous section.
We observe furthermore that the modified modular
transformations involving the iterated period integral
increases the exponent in the bound.

\subsection{Bound for coefficients of modular forms and cusp forms}\label{sec41}
We review in this section the elementary saddle point analysis to determine a rough
bound for modular forms. Since we know the bound is not sharp, we will
not be careful with errors to the bound.
\vspace{.2cm}\\
{\bf Modular forms}\\
First we consider a modular form of weight $k$, $f(\tau)=\sum_{n\ge 0} a(n) q^n$ and $n\in\mathbb{Z}$ and $a(0)\ne 0$. Its transformation property is given by,
\begin{eqnarray}
f\!\left(-\frac{1}{\tau}\right)=C\tau^k f(\tau),
\end{eqnarray}
with $C$ a phase. The $n$-th Fourier coefficient is given by, 
\begin{eqnarray}
a(n) &=& \int_0 ^1 f(\tau)\,e^{-2\pi in\tau}\,d\tau\;, \\ \nn
  |a(n)| & \leq & \int_0 ^1 \left|f\!\left(-\frac{1}{\tau}\right)\,\tau^{-k}e^{-2\pi in\tau}\right|\,d\tau \;.
\end{eqnarray}
We assume that ${\rm Im}(-1/\tau)\gg 0$ at the saddle point, such
that the saddle point can be determined from the first non-vanishing term in the $q$-expansion, $a(0)$. This gives for the saddle point $\tau_s$,
\begin{eqnarray}
-\frac{k}{\tau_s}=2\pi in, \quad \tau_s=\frac{ik}{2\pi n}.
\end{eqnarray}
Thus the assumption is indeed satisfied for sufficiently large $n$.
The upper bound on $a(n)$ is given by,
\begin{eqnarray}
|a(n)|<  c\,n^k\;.
\end{eqnarray}
with for the constant $c=|a(0)|\left(\frac{2\pi e}{k}\right)^k$. This is
clearly not a sharp bound, since we know from the discussion on Eisenstein series that the
bound for modular forms is proportional to $p^{k-1}$ for prime
coefficients $a(p)$.
\vspace{.2cm}\\
{\bf Cusp forms}\\
Now we show that the same method applied to a cusp form
$$S(\tau)=\sum_{n\geq \Delta>0}a(n) q^n$$ of
weight $k$, gives $c\,n^{k/2}$ for some constant $c$ for the bound of
its Fourier coefficients. We assume the following transformation property for $S(\tau)$
\begin{eqnarray}
S\left(-\frac{1}{\tau}\right) &=& C\tau^k S(\tau),
\end{eqnarray}
for a phase $C$. Assuming again that ${\rm Im}(-1/\tau)\gg 0$ at the
saddle point, we
can approximate $S(-1/\tau)$ by its first non-vanishing term
$a(\Delta)\,e^{-2\pi i \Delta/\tau}$. We then have
\begin{eqnarray}
|a(n)| &\leq & \int_{0}^1 \left|f(\tau)e^{-2\pi in\tau}\right|\,
               d\tau\approx \int_{0}^1 \left|\tau^{-k} a(\Delta)\, e^{-2\pi i\Delta/\tau}e^{-2\pi in\tau}\right|\,d\tau.
\end{eqnarray}
Extremization of $-k\log(\tau)-2\pi i \Delta /\tau -2\pi i n\tau$,
gives for the saddle point,
\begin{eqnarray} 
\tau_s &=& \frac{-k\pm \sqrt{k^2-16\pi^2\Delta n}}{4\pi 
           in}=\frac{-k\pm i \sqrt{\delta_n}}{4\pi i n},
\end{eqnarray}
where we introduced $\delta_n=16\pi^2\Delta n-k^2$. In the regime of
interest, large $n$, $\delta_n>0$, and the assumption ${\rm
  Im}(-1/\tau_s)\gg 0$ is satisfied. We have for the magnitude
\be
|\tau_s|^2=\frac{\Delta}{n}.  
\ee

Substitution of $\tau_s$ in the exponents, gives
\be
e^{-2\pi i\Delta/\tau_s}e^{-2\pi in\tau_s}=e^{\mp i \delta_n}.
\ee
This has unit magnitude and thus does not contribute to the magnitude
of the integrand. We therefore find for the bound,
\be
\label{cuspsaddle}
|a(n)|\leq c\,n^{k/2},
\ee
with $c=\Delta^{-k/2}|a(\Delta)|$. This bound is known as the Hecke bound.
We note that this bound is less accurate than Deligne's theorem ie, for a cusp form of weight $k$ we expect the growth of the coefficients to grow maximally as $\sigma_0(n)n^{\frac{k-1}{2}}$, where $\sigma_0(n)$ is the number of divisors of $n$.

\subsection{Bound for coefficients of mock cusp forms}\label{sec42}
Next we proceed with the mock cusp form $S_\mu$ (\ref{defvecS}).
From Eq. (\ref{S3mu}), we have the the following transformation 
\be 
\label{Smutrafo} 
\begin{split}
  S_{0}(\tau) &= \frac{-i\tau^{-3}}{\sqrt{3}}\left(S_0(-1/\tau)+2S_1(-1/\tau)+{\cal J}_0(-1/\tau)\right),\\ 
S_{1}(\tau) &=
\frac{-i\tau^{-3}}{\sqrt{3}}\left(S_0(-1/\tau)-S_1(-1/\tau)+{\cal
    J}_1(-1/\tau)\right),
\end{split}
\ee
where, 
\begin{eqnarray}
{\cal J}_0(\tau) &=& \frac{\sqrt{3}i}{2\sqrt{2}\pi}\sum_{\nu=0}^2\sum_{\alpha=0}^1\int_{0}^{i\infty}\frac{\widehat f_{2,\alpha}(\tau,w)\Theta_{\frac{2\nu+3\alpha}{6}}(3w)}{(-i(w+\tau))^{3/2}}\; dw,\\ \nn
{\cal J}_1(\tau) &=& \frac{\sqrt{3}i}{2\sqrt{2}\pi}\sum_{\nu=0}^2
                     \sum_{\alpha=0}^1 e^{-2\pi i\nu/3} \int_{0}^{i\infty}\frac{\widehat f_{2,\alpha}(\tau,w)\Theta_{\frac{2\nu+3\alpha}{6}}(3w)}{(-i(w+\tau))^{3/2}}\; dw.
\end{eqnarray} 
Similarly to the discussion above, we have
\be
\begin{split}
  |s_\mu(n)|&\leq \int_0^1 \left|S_\mu(\tau)\,e^{-2\pi i n\tau}\right| d\tau.
\end{split}
\ee
We then have to determine the leading contribution among the terms on
the rhs of (\ref{Smutrafo}).  For large ${\rm Im}(-1/\tau)$, the
leading terms of $S_\mu(-1/\tau)$ follow from
(\ref{Smucoeffs}). These are exponentially decreasing. To compare
these with ${\cal J}_\mu$, we recall Lemma 3.1 in
\cite{Bringmann:2010sd}, (see also \cite{Bringmann2007})
\begin{eqnarray} 
\int_{0}^{\infty}\frac{\Theta_{\alpha/2}(iz)}{(z+x)^{3/2}}
  dz=\frac{2}{\sqrt{x}}\delta_{\alpha,0}+ O(x^{-3/2}),
\end{eqnarray} 
with the leading term coming from the constant term of
$\Theta_{0}$. As a result, we have
\be
\begin{split}
\mathcal{J}_\mu(\tau)=-\frac{\sqrt{3}i}{2\sqrt{2}\pi} \frac{1}{4}\frac{2}{\sqrt{-i\tau}}+O(\tau^{-3/2}).
\end{split}
\ee
Thus the contribution from $\mathcal{J}_\mu(-1/\tau)$ provides the
leading term to the rhs in (\ref{Smutrafo}). We thus have for the coefficients
\be
|s_\mu(n)|\leq \frac{1}{4\sqrt{2}\pi}\int_0^1 \left| \tau^{-5/2}
  e^{-2\pi i n\tau} \right| d\tau.
\ee
Using the saddle point method, we then arrive at
\be
|s_\mu(n)|\leq c\,n^{5/2},
\ee
with $c=\frac{1}{4\sqrt{2}\pi} (4\pi e/5)^{5/2}$. This upperbound
agrees with the numerical experiments, in the sense that the numerics suggest a growth
proportional to $n^{3/2}$, which is clearly much smaller than
$n^{5/2}$. We note that for a classical cusp form terms like ${\cal J}_\mu$
are absent in the modular transformation, and the saddle point method
(\ref{cuspsaddle}) gives $n^{3/2}$ for the growth.

\subsection{Average growth for theta series}
\label{sec53}
If more is known about the arithmetic nature of the coefficients,
sharper bounds can often be obtained than the saddle point
method. This is for example the case of Eisenstein series and cusp
forms. Often the modular forms can also be expressed as lattice sum or
theta series. 

We give a heuristic argument (probably well-known to many) that the
leading term of the average growth of the coefficients of a theta series equals that of
Eisenstein series. This should of course be the case since all theta
series which transform as modular forms can be expressed as linear
combinations of Eisenstein series. To make the argument, let us
consider a $d$-dimensional positive definite lattice $L$ with integral
quadratic form $Q$. The associated theta series $\Theta_L(\tau)$ and
coefficients $d(n)$ are defined through,
\be
\begin{split}
\Theta_L(\tau)&=\sum_{k\in L}q^{Q(k)/2}\\
&=\sum_{n\geq 0} d(n)\,q^n.
\end{split}
\ee
We also introduce the cumulative sum $D(N)$ as, 
\be
D(N)=\sum_{0\leq n\leq N} d(n),
\ee
such that
\be
d(N)=D(N)-D(N-1).
\ee

For a theta series, $D(N)$ is a count of lattice points and thus
scales as the volume of the domain in $L$, whose lattice points are
enumerated by $D(N)$. The volume scales on average as $|k|^d=|Q(k)|^{d/2}=N^{d/2}$, such
that $d(N)=D(N)-D(N-1)$ scales as $N^{(d-2)/2}$. Since the weight $k$ of
$\Theta_L$ is $k=d/2$, we find for the leading term
\be
d(n) \approx C\,n^{k-1}. 
\ee  
 
We can use the same approximation for indefinite theta series, which
involve a sum over a positive definite cone in an indefinite
lattice. The class number generating function $G_{2,\mu}$ can be
expressed in this form. Moreover, the expression of Kool \cite{Kool_2014} for $f_{3,1}$ are of this
form. This matches our observation that the coefficients of
$f_{3,\mu}$ grow as $n^{2}$, similar to the Eisenstein series of
weight 3. This rough estimate is for the coefficients of $f_{3,\mu}$ is even sharper than
the bound for the coefficients of $S_\mu$ using the saddle point
method.

\section*{Acknowledgements}
The majority of this work was carried out while AC was a postdoctoral
fellow in the School of Mathematics, Trinity College Dublin. During
this time, AC and JM were supported by the Laureate Award 15175
“Modularity in Quantum Field Theory and Gravity” of the Irish Research
Council. JM is also supported by the Ambrose Monell Foundation. AC is presently funded by fellowship from DIAS. We thank Prof. Werner Nahm for useful discussions. We also thank the participants of ``Modular Forms in Number Theory and Beyond", Bielefeld especially Olivia Beckwith, Jan Vonk, Caner Nazaroglu and Michael Mertens for useful insights.

\appendix
\section{Explicit expressions for $f_{N,\mu}$ for $N=2,3$}
The explicit expressions for the class number generating functions $G_\mu$ as $q$-series
were given in \cite{Bringmann:2010sd}, as:
\begin{eqnarray}
\label{f2021series}
G_{0}(\tau) &=& -\frac{1}{2\Theta_0(\tau+1/2)}\sum_{n\in\mathbb{Z}}\frac{n(-1)^n q^{n^2}}{1+q^{2n}}-\frac{1}{12}\Theta_0^3(\tau),\\ \nn
G_{1}(\tau) &=& -\frac{q^{-1/4}}{2\Theta_0(\tau)}\sum_{n\in\mathbb{Z}}\frac{(2n-1)q^{n^2}}{(1-q^{2n-1})}+\frac{1}{6}\Theta_{1}^3(\tau),
\end{eqnarray}
where,
\begin{eqnarray}
\Theta_{j}(\tau) &=& \sum_{n\in\mathbb{Z}}q^{\frac{(2n+j)^2}{4}}.
\end{eqnarray} 
The explicit expressions for $f_{3,\mu}$ as expansions in $q$ series were given in \cite{Manschot:2017xcr}, which are quoted as follows:
\begin{eqnarray}
b_{3,0}(\tau)f_{3,0}(\tau) &=& \frac{13}{240}+\frac{E_2(\tau)}{24}+\frac{E_2(\tau)^2}{72}+\frac{E_4(\tau)}{720}\\ \nn
&&-\frac{9}{2}\sum_{k\in\mathbb{Z}}k^2 q^{3k^2}+\sum_{k_1,k_2\in\mathbb{Z}}(k_1+2k_2)^2 q^{k_1^2+k_2^2+k_1k_2}\\ \nn
&&+ \sum_{\begin{smallmatrix}k\in\mathbb{Z}\\k\ne 0\end{smallmatrix}} S_{1,0}(k,q)q^{3k^2}\\ \nn
&&+ \sum_{\begin{smallmatrix}k_1,k_2\in\mathbb{Z}\\2k_1+k_2\ne 0\\ k_2-k_1\ne 0 \end{smallmatrix}} S_{2}(2k_1+k_2,k_2-k_1,q)q^{k_1^2+k_2^2+k_1k_2+2k_1+k_2}
\end{eqnarray}
\begin{eqnarray} 
b_{3,0}(\tau)f_{3,1}(\tau) &=& 
                               \sum_{\begin{smallmatrix}k\in\mathbb{Z} \end{smallmatrix}}
                               S_{1,1}(k,q)q^{3k^2-1/3}\\ \nn
&&+ \sum_{\begin{smallmatrix}k_1,k_2\in\mathbb{Z}\\2k_1+k_2\ne 1\\ k_2-k_1\ne 0 \end{smallmatrix}} S_{2}(2k_1+k_2-1,k_2-k_1,q)q^{k_1^2+k_2^2+k_1k_2+2k_1+k_2-1/3}
\end{eqnarray}
where we have,
\begin{eqnarray}\nn
S_{1,\mu}(k,q) &=& \frac{(E_2(\tau)-1)(k-\mu+1)}{2(1-q^{3k-\mu})}+\frac{9(k-\mu)^2+33(k-\mu)+31-E_2(\tau)}{2(1-q^{3k-\mu})^2}\\
&&-\frac{15(k-\mu)+34}{(1-q^{k-\mu})^3}+\frac{19}{(1-q^{3k-\mu})^4}\\ \nn
S_{2}(a,b,q) &=& \frac{4q^b}{(1-q^a)(1-q^b)^3}+\frac{4q^a}{(1-q^b)(1-q^a)^3}+\frac{4}{(1-q^a)^2(1-q^b)^2}\\ \nn
&&-\frac{2q^b(a+b+1)}{(1-q^a)(1-q^b)^2}-\frac{2q^a(a+b+1)}{(1-q^b)(1-q^a)^2}+\frac{(a+b-2)^2-8}{(1-q^a)(1-q^b)}
\end{eqnarray}
The blow-up formula provides relations among different representations
for the $q$-series of $f_{3,\mu}$ \cite{Bringmann_2016}.

The Eisenstein series $E_2$ and $E_4$ are given by,
\begin{eqnarray}
E_2(\tau) &=& 1-24\sum_{n=1}^{\infty}\sigma_1(n)q^n,\quad E_4(\tau) = 1+240\sum_{n=1}^{\infty}\sigma_3(n)q^n
\end{eqnarray}
for $\sigma_k(n)=\sum_{d|n}d^k$.

\providecommand{\href}[2]{#2}\begingroup\raggedright\endgroup


\end{document}